\documentclass[%
 reprint,
superscriptaddress,
 amsmath,amssymb,
,prl
]{revtex4-1}
\usepackage{xcolor}

\usepackage{graphicx}
\usepackage{dcolumn}
\usepackage{bm}
\usepackage{texshade}

\begin{document}

\preprint{APS/123-QED}

\title{Consistent treatment of hydrophobicity in protein lattice models accounts 
for cold denaturation}

\author{Erik van Dijk}
 \email{e.van.dijk@vu.nl}
\affiliation{
 Department of Chemistry, University of Cambridge, Cambridge CB2 1EW, United Kingdom
}%
\affiliation{Centre for Integrative Bioinformatics (IBIVU), VU University Amsterdam, The Netherlands}
\author{Patrick Varilly}%
\affiliation{
 Department of Chemistry, University of Cambridge, Cambridge CB2 1EW, United Kingdom
}%
 \email{patvarilly@gmail.com }

\author{Tuomas Knowles}
\affiliation{
 Department of Chemistry, University of Cambridge, Cambridge CB2 1EW, United Kingdom
}%
\email{tpjk2@cam.ac.uk}
\author{Daan Frenkel}
\affiliation{
 Department of Chemistry, University of Cambridge, Cambridge CB2 1EW, United Kingdom
}%
 \email{df246@cam.ac.uk}

\author{Sanne Abeln}
\email{s.abeln@vu.nl}
\affiliation{Centre for Integrative Bioinformatics (IBIVU), VU University Amsterdam, The Netherlands}

\date{\today}

\begin{abstract}

The hydrophobic effect stabilizes the native structure of proteins by minimizing the unfavourable interactions between hydrophobic residues and water through the formation of a hydrophobic core. Here we include the entropic and enthalpic contributions of the hydrophobic effect explicitly in an implicit solvent model. This allows us to capture two important effects: a length-scale dependence and a temperature dependence for the solvation of a hydrophobic particle. This consistent treatment of the hydrophobic effect explains cold denaturation and heat capacity measurements of solvated proteins.

\end{abstract}


\maketitle

 The 
stability of the native state of most proteins is typically dominated by 
interactions between amino acids and through the hydrophobic effect. The direct 
amino-acid interactions can be attributed to van der Waals and electrostatic 
forces that are mainly enthalpic in nature. By contrast, hydrophobicity is an 
interaction emerging from the collective behaviour of the solvent and the side 
chains, and is entropy dominated~\cite{Lum1999,Huang2000,Kim1992,Widom2003,Chandler2005} 
at room temperature for small solutes. The enthalpic amino acid interactions 
remain relatively constant over the temperature range of interest, while the 
magnitude of the hydrophobic effect changes with temperature~\cite{Widom2003,Huang2000}.

In principle, all-atom simulations could be used to disentangle the role of entropy and enthalpy in protein folding. However, fully atomistic simulations are neither simple, nor cheap - in fact, at present, such simulations are only feasible to study the folding of relatively small proteins. Moreover, a numerical study of the stability of various protein structures would require simulations over a range of temperatures. Earlier studies~\cite{Dias2008, Romero-VargasCastrillon2012, Bianco2015} have shown that a temperature-dependent hydrophobic collapse (rather than folding) can be observed in a strongly coarse-grained model for small, two-dimensional protein chains. Three-dimensional models have shown similar results for homopolymers \cite{Bianco2015} and peptides \cite{Mitsutake2004}. However, these models do not fully capture folding specificity for proteins.

In this work we present an extension of the classic protein lattice model first
introduced in Ref.~\cite{Sali1994}. The classic model correctly reproduces
the ability of proteins to fold into a unique native structure and it
exhibits denaturation upon heating due to chain entropy alone. Interactions
between amino acids are estimated through the frequency of occurrence of
close contacts in experimental protein crystal structures \cite{Miyazawa1985}. In the Miyazawa and Jernigan (MJ) potential, the interactions are strictly speaking free energies that have both entropic and enthalpic components.  However, in most coarse-grained simulations these effective potentials are treated as temperature-independent enthalpies. Therefore, they do not model the temperature dependence of the interactions correctly.

In order to model the temperature dependence of the hydrophobic effect, we use an extension of the MJ potential that includes specific solvent-amino acid interaction terms~\cite{Abeln2011}. The derived potential is based on a representative subset of the Protein Database \cite{Griep2010}. The hydrophobic effect is volume dominated at small length scales and surface dominated at large length scales. Our model consistently treats this length-scale dependence by dynamically classifying residues into three categories: buried, protein surface, and fully solvated. This categorization allows us to capture the length-scale dependence of the hydrophobic effect according to Lum-Chandler-Weeks (LCW) theory \cite{Lum1999,Andersen1971}. Our model aims to reproduce the variation in temperature 
dependence for different length scales of hydrophobic solutes using these implicit solvent terms. 

\begin{figure}
\centering
\includegraphics[width=\linewidth]{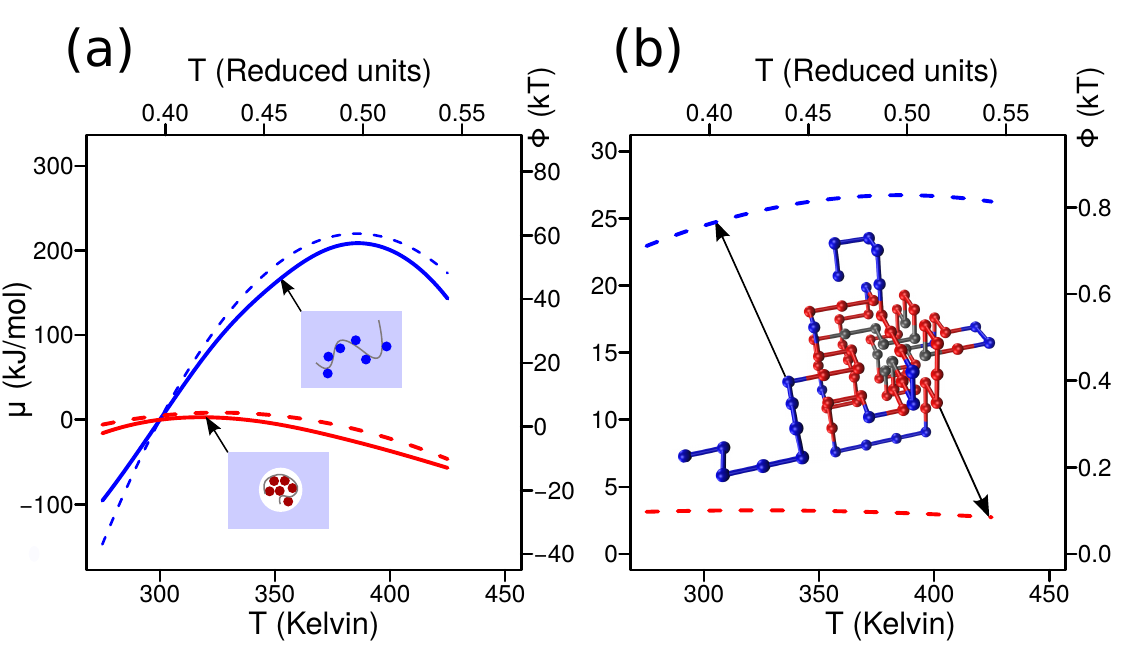}
\caption{Comparison between lattice model and LCW-theory for a poly-phenylalanine hydrophobic chain.
(a) The chemical potential for a fully extended chain as a function of temperature (blue lines), and the chemical potential for a compacted chain, which we approximate as a 10 \AA\ sphere (red lines). The dashed lines indicate the approximation made by our lattice model, while the solid lines indicate the theoretical predictions from LCW-theory \cite{Lum1999,Huang2000}.
(b) The distinction between surface and fully solvated residues in our model. The blue line shows the potential for the fully solvated residues (corresponding to the residues colored blue), and the red line shows the surface potential (corresponding to the residues colored red). \label{fig:Model}
}
\end{figure}

For each residue category, the hydrophobe-water interaction is estimated by a second-order Taylor approximation to the free energy of
transfer of hydrophobic particles from an oily environment to water (see Figure~\ref{fig:Model}). We use this model to investigate three effects that are often associated with the temperature dependence of the hydrophobic effect: Firstly, denaturation upon cooling, or `cold denaturation'. Cold denaturation conflicts with the classical view of an entropically favourable state and an enthalpically favourable native state. Secondly, the structural characteristics of the cold denatured state.  Thirdly, the temperature dependence of the heat capacity. Using Differential Scanning Calorimetry (DSC) \cite{Hallerbach1999} the heat capacity of the system can be calculated as $C_P = 
\left(\frac{d Q}{d T}\right)_{P,N}$. The heat capacity of the system is commonly used as a well defined experimental observable to characterize the thermodynamics of the folding transition.

We simulate a
protein consisting of $80$~residues with Monte Carlo sampling using a classic lattice model to investigate the effect of the entropic contribution of the hydrophobic
potential. To
model the effective potential for hydrophobe--water interactions, we
introduce the following temperature-dependent term for the surface residues ($s$) and the fully hydrated residues ($h$):
\begin{equation}
F_{\text{hydr}}=-\alpha_s N_{\text{s}} (T-T_{0,s})^2 -\alpha_h N_{\text{h}} (T-T_{0,h})^2
\label{eq:TempDepPotential}
\end{equation}
describing second order approximations to the theory of the hydrophobic 
effect~\cite{Huang2000,Chandler2005} for both groups.
Here, $N_{\text{s}}$ is the number of hydrophobic residues on the surface,
$N_{\text{h}}$ is the number of hydrophobic residues that are fully hydrated and
$T$ is the temperature in reduced units. The temperature dependence of the fully hydrated ($\alpha_h$, $T_{0,h}$) and surface ($\alpha_s$, $T_{0,s}$) residues are set using Ref.~\cite{Chandler2005} (see 
Figure~\ref{fig:Model}(a)).  In our lattice model, we define a residue that is 
fully hydrated as having at least four sides exposed and for a residue that is 
partially solvated as having at least one, and no more than three sides exposed to the 
solvent. Fitting the expression in eqn.~\eqref{eq:TempDepPotential} to 
the results~\cite{Huang2000} from LCW theory yields $\alpha_s=3.0$ and $T_{0, s}=0.41$ for 
the surface term and $\alpha_h = 7.0$ and $T_{0, h}=0.49$ for the volume 
solvation term. This assumes that, for the temperature dependence, all amino acids have the same size, while in practice, the volume of amino acids can vary from 75 to 240 \AA \cite{Mishra1984}. To test the sensitivity of our model to this assumption, we performed simulations with three different potentials: a temperature independent potential ($\alpha_s=\alpha_h=0$, a temperature dependent potential (parameters given above) and a strongly temperature dependent potential, corresponding to amino acids that are 15\% larger ($\alpha_s=4.5$ and $\alpha_h=11.5$) (Derivation shown in SI sections ``Derivation temperature dependent potential'' and ``Approximation of hydrophobicity parameters'').

First, we probe the folding specificity of this model. The lattice model we use here is sequence dependent. In other words, random sequences will typically not fold into a stable structure, whereas designed sequences do so with a high specificity  \cite{Shakhnovich1991,Shakhnovich1993a,Shakhnovich1993b,Shakhnovich1994,Coluzza2003,Abeln2008,Abeln2011}.

 The number of native contacts 
($N_{\text{int}}$) is used as an order parameter for the specificity of protein 
folding. We define a protein to be folded when $N_{\text{int}}>75$. The fraction 
of the simulation spent in this folded state is defined as $P_{\text{Fold}}$. 
For this work umbrella sampling~\cite{Grossfield2003}  alone is sufficient to 
sample the configurational space of interest. Figure~\ref{fig:EnergyLandscape}(a) shows that for all potentials the protein folds ($P_\text{Fold}>0.5$) at 
intermediate temperatures and denatures ($P_\text{Fold}<0.5$) at high 
temperatures. This is consistent with the view of the high-entropy denatured 
state caused by the chain entropy. For a well-designed protein, the stability of 
the protein simulated with a temperature-independent potential is a strictly 
decreasing function of the temperature, since the native state is optimized to 
be the lowest enthalpy state.

\begin{figure*}
\includegraphics[scale=1]{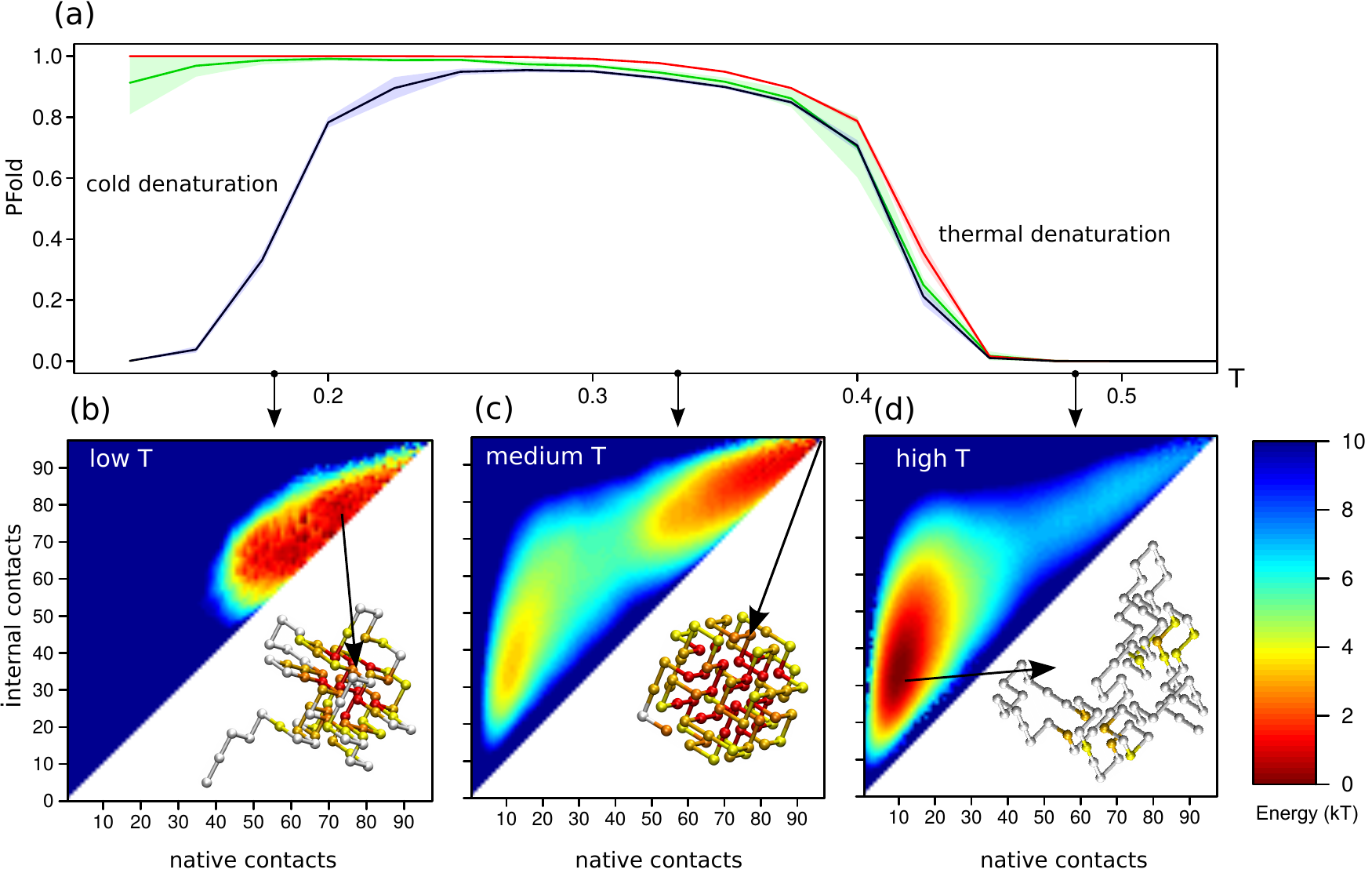}
\caption{Temperature-dependent folding stability and structure. The folded state has 97 native contacts.\\ (a)
The probability for the model protein to be in the folded state versus
temperature, with $\alpha=0$ (red), the temperature dependent potential (green) and a simulation where the temperature dependence is multiplied by 1.5 (black). The 95\% confidence interval is indicated by the shaded area.
(b)--(d) Free energy landscapes for the number of native contacts
($N_{\text{int}}$) and all internal contacts ($C_{\text{int}}$) for
the simulations with a strong temperature dependence at (b) low temperature ($T=0.175$), (c) intermediate
temperature ($T=0.375$), and (d) high temperature ($T=0.475$). For
the strong temperature dependent potential, the protein denatures at low temperatures, with many exposed
hydrophobic amino acids. However, this denatured structure is a lot more
compact than the heat denatured protein, and there are less native contacts present. At intermediate temperatures
the protein has the highest stability in its folded configuration
(indicated by the arrow) where $N_{\text{int}}=C_{\text{int}}=97$. At
high temperatures the protein makes only transient contacts.}
\label{fig:EnergyLandscape}
\end{figure*}

Only the strongly temperature-dependent potential reproduces cold denaturation as well as heat-induced denaturation, see black curve in Figure~\ref{fig:EnergyLandscape}(a). A very similar folding curve has been observed experimentally for a mutant of cold shock protein Csp \cite{Szyperski2006}. The simulated configurational ensemble of the folded state also includes a small fraction denatured states ($0.20 <T <0.37 $)  as observed in the experiment. Note that Csp, like most proteins, does not show cold denaturation above the freezing point of water. However, statistical investigation has shown that the temperature has a measurable influence on the propensity of hydrophobic amino acids to be buried \cite{vanDijk2015}.  This is similar to our observation that proteins become less stable at lower temperatures, but do not denature, for a lower value of the temperature dependence. (Figure~\ref{fig:EnergyLandscape}, green line).

The structural characteristics of the model were investigated by exploring
the free energy landscapes of native contacts ($N_{\text{int}}$) and internal contacts between
residues ($C_{\text{int}}$); the latter are used as a measure of
compactness. At T=0.375, slightly below the transition temperature (T=0.42),
two distinct states can be observed, one where the protein is specifically
folded ($N_{\text{int}} > 75$), and one in which the protein is mostly
unstructured ($N_{\text{int}} < 25$), with a clear barrier separating the
two states (Figure~\ref{fig:EnergyLandscape}(c)). Note that the sequence
has been designed to fold in this exact structure with 97 native contacts
(see Methods). 

Comparing Figures~\ref{fig:EnergyLandscape}(b) and (d)
it becomes apparent that the cold denatured state has more residual structure than the heat denatured state. The cold denatured configurational ensemble at T=0.125 shows
a structure that is compact with approximately two thirds of the native
contacts present, similar to experimental NMR observations of pressure-assisted cold denaturation \cite{Vajpai2013}, urea-assisted cold denaturation \cite{Wong1996} and cold denaturation for a protein that was destabilized by a mutation \cite{Shan2010}. Note that for some disordered proteins the radius of gyration decreases as the temperature increases \cite{Wuttke2014,Privalov1993}. This is most likely due to interactions involving charged residues, which play a larger role in disordered proteins. Notably, the $\lambda$-repressor, which is the most hydrophobic protein in the dataset investigated in Ref. \cite{Wuttke2014}, does show a re-expansion at temperatures higher than 319 K \cite{Wuttke2014}.

In addition to the structural characteristics, our model allows us to investigate the role of the hydrophobic effect in the thermodynamics of protein folding.  We start by investigating the heat capacity of folding. Note that the simulations are performed at constant volume, while the experiments are done at constant pressure. However, the difference is neglible for the system in  consideration due to the low compressibility of water (See SI for more detail). In order to calculate the heat capacity ($C_V$ in our model) for the temperature-dependent potential, we need to separate the expected enthalpy $<E>$ from the entropic part of the hydrophobic
potential, $F_{\text{hydr}}$ (see SI). In a finite system, a phase transition is usually characterized by a sharp peak in the heat capacity that can be observed experimentally \cite{Privalov1989}. For the temperature-independent potential we observe only a single peak at the folding transition (Figure S3(a)). In contrast, the heat capacity of the temperature dependent potential shows two peaks, one for cold induced denaturation and one for heat induced denaturation (Figure S3(b)).
\begin{figure}
\includegraphics{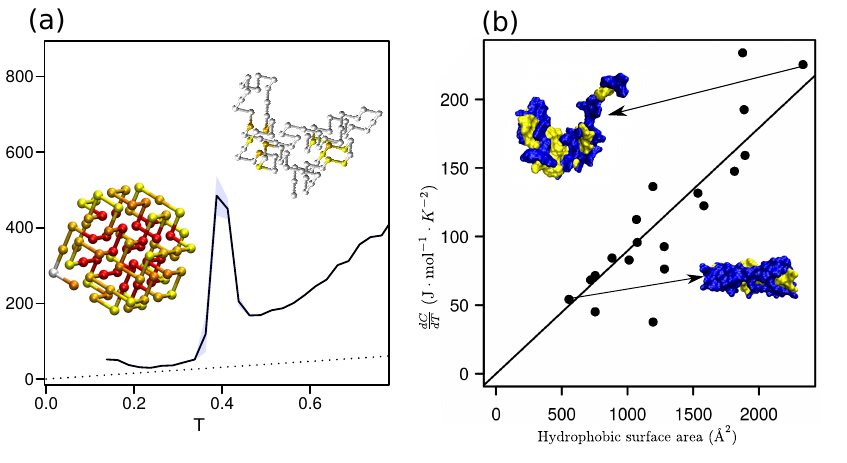}
\caption{Heat capacity versus temperature
(a) The heat capacity for the fitted potential as calculated by $C_V=\frac{dE}{dT}$, corresponding to the green line in Figure \ref{fig:EnergyLandscape}(a), shows a linear increase in the heat capacity with respect to temperature. Our model suggests that the slope of this baseline, which is a lower bound for the heat capacity, should correlate with the amount of hydrophobic surface area. (b) The slope of the heat capacity $\frac{dC}{dT}$, shows a strong correlation with the exposed hydrophobic surface area in real protein structure as predicted by our model. Two indicative protein structures, with pdb-codes 1J46 and 2ZTA are shown with the hydrophobic and hydrophilic amino acids colored yellow and blue respectively.}
\label{fig:HeatCapacity1}
\end{figure}
Another interesting observation is a linear temperature dependence of the heat 
capacity in the temperature range where no phase transition is occurring 
(Figure~\ref{fig:HeatCapacity1}(a)).  The slope or the linear increase in the baseline of the heat capacity has been investigated in Refs. \cite{Privalov2007,Munoz2004,Farber2010,Bruscolini2011,Johnson2011}. In the context of
the current model we can understand the linear $T$-dependence of the heat capacity in terms of the exposed hydrophobic groups. Assuming a constant hydrophobe-water contact area and neglecting entropic contributions other than the hydrophobe-water contact area, we can derive a simple lower bound for the heat capacity (see SI for derivation):
\begin{equation}
C_V(T) = (2 \alpha_s N_{\text{s}} + 2 \alpha_h N_\text{h}) T
\label{eq:LinearHeatCapacity}
\end{equation}
where $N_{\text{s}}$ is the number of hydrophobic amino acids that are at the surface, and $N_\text{h}$ the number of hydrophobic amino acids that are fully hydrated. This simple calculation yields a lower bound for the heat
capacity in regions where no folding transition occurs, as indicated by the dotted lines
in Figure \ref{fig:HeatCapacity1}(a). Here we estimate $N_\text{s}$ and $N_\text{h}$ for the 
folded regime from the number
of exposed hydrophobes in the native structure
($N_\text{s}=13$, $N_\text{h}=0$). The difference between the lower bound shown in Figure \ref{fig:HeatCapacity1}(a) and the simulated results is likely due to the chain entropy in the native ensemble. This is supported by the results from the temperature independent potential (see also Figure S3(a)).

Equation \eqref{eq:LinearHeatCapacity} has additional consequences that can be
verified experimentally. Initially, as a consistency check, it is easy to see that from eqn. \eqref{eq:LinearHeatCapacity} we can recover the well known relationship between the
change in heat capacity, $\Delta C_V$ and the change in hydrophobic surface area upon folding at a given temperature $T$, eg. Refs \cite{Spolar1989,Anslyn2006}. We can go further, however, and probe the
derivative of the heat capacity with respect to the temperature
($\frac{dC_V}{dT}$), as shown by the dotted lines in Figure~\ref{fig:HeatCapacity1}(a). Our analysis in eqn. \eqref{eq:LinearHeatCapacity} predicts that this slope
itself is proportional to the exposed hydrophobic surface area (corresponding to $N_h$ in the model) in a given state.

To test this prediction we compared the heat capacity slopes for folded proteins
tabulated in Ref.~\cite{Privalov2007} with the level of exposed
hydrophobic surface area in the corresponding folded structures obtained with \cite{Kabsch1983}. We find a
strong correlation ($R^2=0.77$) between the slope of the heat capacity and the
exposed hydrophobic surface area (see Figure~\ref{fig:HeatCapacity1}(b)), further
supporting the temperature dependence of the hydrophobic effect as a key
mechanism underlying the linear increase of the heat capacity. 

Previously, a higher slope for DNA-binding proteins has been observed when compared to globular proteins of the same size \cite[see also SI Figure 2]{Gomez1995,Privalov2007,Naganathan2011} was rationalized through the flexibility of DNA-binding domains \cite{Privalov2007,Naganathan2011}. Our work suggests that the increased slope of the heat capacity may be explained solely by a higher exposed hydrophobic surface area. Hence, the strong correlation between the slope of the heat capacity and the exposed hydrophobic surface area for all proteins may be explained by a consistent treatment of hydrophobicity alone (see Figure S2). This does not, however, preclude a possible correlation between the flexibility and the amount of exposed surface area.

To conclude, we have presented an extension for a coarse-grained lattice model by including a temperature-dependent hydrophobic term in the interaction potential. The combination of the coarse-grained steric model and the potential ensures appropriate contributions of the solvent-residue and internal contacts. This allows us to separate chain entropy from the solvent entropy. An effective quadratic potential is applied to account for the temperature dependence of the hydrophobic effect. Simulating the model, we observe cold denaturation for realistic parameter settings, suggesting that the hydrophobic effect is the key component in cold denaturation. In addition, we find that the simulated cold denatured state is more compact than the heat denatured state; this is in agreement with experimental observations. Moreover, the model is able to reproduce the characteristic experimental heat capacity curves for protein folding. 

Starting from the temperature dependent potential, we derive a simple relation that approximates the heat capacity baseline of the native state of the protein. This relation is tested for a set of real proteins, where we do indeed find a correlation between the hydrophobic surface area and the slope of the heat capacity. Hence, our model seems to make accurate predictions for thermodynamic behaviour of real proteins. Additionally, the developed relation can potentially be used to calculate an accurate baseline for proteins with a known structure.


 The interaction potential is based on a representative set of the Protein Database, PDB-25 \cite{Griep2010}, and holds some biases towards rigid and soluble proteins, since they are easier to crystallize (see SI for further discussion).  The cubic lattice model also has some limitations: secondary structure can not be modelled explicitly, nor is there sufficient molecular detail to predict the true fold of a protein sequence. However, we stress that the use of our coarse-grained model is not only motivated by considerations of computational cost. Rather, the use of simple, coarse-grained models allows us to reveal the minimal physical ingredients that a model needs to account for cold denaturation.

The temperature-dependent potential developed here is applicable to other (off-lattice) coarse-grained models with implicit solvent-side chain interactions, e.g. Refs. \cite{Hoang2004,Auer2008,Coluzza2011}. The protocol to calculate the heat capacity from a temperature-dependent potential, can be applied to any effective potential with a closed form expression that is continuous and differentiable with respect to $\beta=\frac{1}{k_B T}$. Furthermore, the addition to the lattice model itself will enable investigation of temperature dependence of protein aggregation building on previous studies \cite{Abeln2008,Ni2013,Abeln2014}



\bibliography{MyCollection08212014}

\section*{Supplementary Material \--- Consistent treatment of hydrophobicity in protein lattice models accounts 
for cold denaturation}


\subsection{Monte Carlo simulations}
The simulations were run using the Metropolis-Monte Carlo algorithm. This method consists of making a random trial move, evaluating the Hamiltonian and accepting using the Boltzmann criterium:

\begin{equation}
P_{Acc}=\max(1,\exp(-\mathcal{H}/k_B T))
\end{equation}
where $\mathcal{H}$ is the Hamiltonian, $k_B$ is the Boltzmann constant, and T is the temperature. The complete Hamiltonian of the system is obtained as the sum of the base pair potential and the temperature-dependent correction $F_{\text{hydr}}$ and  (defined in eqns.~\eqref{eq:EBase} and~\eqref{eq:Fhydr}  respectively):

\begin{equation}
\mathcal{H}(T,\vec{r})=E_{\text{base}} +F_{\text{hydr}}
\label{eq:Hamiltonian}
\end{equation}

We calculate the interactions with the solvent separately for surface terms and fully solvated residues

For $E_\text{base}$, we use a standard cubic lattice model, first introduced in Ref.~\cite{Sali1994} (Figure 1). A more detailed description of the model used here can be found in Ref.~\cite{Abeln2011}. In this model, the potential energy $E_{\text{base}}$ of the amino acid interactions can be calculated as follows:

\begin{equation}
\begin{split}
\label{eq:EBase}
E_{\text{base}}=\frac{1}{2}\sum^{N}_{i}\sum^{N}_{j}{\epsilon_{a(i),a(j)} \cdot C_{ij}}+\sum^{N}_{i}{\epsilon_{a(i),w}{C_{\text{wi}}}} +\\ \sum^{N}_{i}{\omega_s \epsilon_{a(i),w}{C_{\text{si}}}}
\end{split}
\end{equation}
Here, $N$ is the number of amino acids, $\epsilon_{a(i),a(j)}$ represents the interaction of amino acid $a(i)$ and $a(j)$. $C_{ij}$ is an adjacency matrix, and is 1 if residues $i$ and $j$ are neighbours in the lattice. A residue can be categorized as buried, ($C_\text{wi}=0$ and $C_\text{si}=0$), on the surface ($C_\text{wi}=0$ and $C_\text{si}=1$), or fully solvated ($C_\text{wi}=1$ and $C_\text{si}=0$). We define a residue to be buried if it has zero contacts with water, on the surface if it has 1-3 contacts with water, and fully solvated if it has 4 or more contacts with water. The weigth $\omega_s$ accounts for the lower interaction energy for a hydrophobic residue at the surface.

\subsection{Derivation temperature dependent potential}

The MJ matrix represents effective interaction free energies. This means it should contain temperature-dependent effects. While van der Waals, electrostatic and other interactions are mostly enthalpic, the hydrophobic interactions are partly entropic and partly enthalpic in nature. Thus, solvent exposed hydrophobic amino acids exhibit a temperature dependence in their excess chemical potential of solvation which is characteristic of small hydrophobes. For small solutes with a radius $r \lesssim\;0.5\mathrm{nm}$, their free energy of solvation is proportional to the volume of the solutes and they can be
accommodated without breaking of hydrogen bonds in the water. In this scenario,
when the hydrophobic amino acids in the model protein are fully solvated, $N_h$
is equal to the total number of hydrophobic amino acids.
When the hydrophobic amino acids cluster at the
core of the protein as it folds, $N_{s}$ will describe the number of surface
exposed amino acids and therefore in this regime the solvation energy emerges as
a surface energy as predicted by theory~\cite{Andersen1971,Chandler2005}. Indeed, in this regime it is
impossible to accommodate the cluster of hydrophobes while maintaining a
complete hydrogen bonding network in the surrounding water and hence there is a
driving force for water molecules to vacate the hydrophobic cluster and an
interface is formed between the bulk water and the water-depleted interior of
the model protein. The energy cost for solvating the cluster of hydrophobes is
therefore dependent on the surface rather than the volume of the solute, as well
as on the surface tension of the air/water interface.

 This means that to correctly model the free energies as a function of temperature, a correction needs to be applied to the interaction columns. For the temperature-dependent potential developed here, we use a quadratic potential:

\begin{equation}
F_{\text{hydr}}=-\alpha_s  N_{\text{s}} (T-T_{0,s})^2 -\alpha_h N_{\text{h}} (T-T_{0,h})^2
\label{eq:Fhydr}
\end{equation}
$T_0$ and $\alpha$ are constants. The subscript $s$ and $h$ indicate the surface and fully hydrated terms. The complete Hamiltonian of the system is obtained as the sum of the temperature-dependent correction $F_{\text{hydr}}$ and the base pair potential (defined in eqns.~\eqref{eq:Hamiltonian},~\eqref{eq:EBase} and~\eqref{eq:Fhydr}) respectively.

\subsection*{Approximation of hydrophobicity parameters}

To relate our model to theoretical calculations, we introduce a new function, $\phi(T)$, which corresponds to chemical potential of solvation for a single residue. For a single fully hydrated residue, $\phi_h(T) = \mu$.  This means that for any single configuration of a lattice protein, we can calculate the chemical potential of solvation as $\mu_{\text{lattice}}=N_s \phi_s(T) + N_h \phi_h(T)$.

We calculate $\phi(T)$ separately for surface terms and fully solvated residues. For a fully solvated residue, the interaction is approximated by a quadratic fit to the potential for a 3.3 \AA\ particle to data in Ref.~\cite{Huang2000}. Since the model is run in reduced units, this potential needs to be rescaled to be consistent with the statistical potential. $T_{\text{amb}}$, is taken to be 300K, where proteins are stable and their structure is being determined. $\Phi_h(T_{\text{amb}})$ is the corresponding interaction of the hydrophobic particle with water in the statistical potential. This yields a reference point for the potential, allowing the potential to be rescaled to the interactions in the lattice model. We write the reduced temperature, $T^*$, as: $T^*=kT$. In our model, $T^*_{\text{amb}}=0.4$, which corresponds to 300 K. Solving for $k$ yields $k=1.3 \cdot 10^{-3} K^{-1}$. Unless otherwise specified, temperatures in this paper are given in reduced units.

The weight of the surface term, $\omega_s$, used to rescale the surface potential using:
\begin{equation}
\label{eq:SurfaceHydrated}
 \Phi_s(T_\text{Amb})=\omega_s \Phi_h(T_\text{amb})
\end{equation}

 is fitted by considering a fully hydrophobic chain in the native structure of the lattice model, counting the number of surface terms, and distributing the solvation free energy over the surface in our model. The native structure consists of 24 buried residues, 1 fully solvated residue and 55 surface residues. We use a sphere with a radius of 10 \AA\ to approximate the size of a protein.

The ratio of the surface terms with respect to the volume terms is fitted such that the ratio of the transfer free energy, $\mu$, of the 10 \AA\ particle to the 3.3 \AA\ is the same as in the theoretical model. The values resulting from this procedure: $\alpha_s=3.0$, $T_0=0.41$ and $\omega_s=0.41$ for the surface terms and $\alpha_h = 7.0$ and $T_0=0.49$ for the fully solvated terms. 
\begin{equation}
\label{eq:LCWFit}
\begin{split}
\frac{ \mu_{_\text{LCW}}(r_{\text{protein}},T_{amb})}{\mu_{_\text{LCW}}(r_{\text{residue}},T_{amb})}=\\ 
\frac{N_s \omega_s \phi_{s,\text{PHE}}(T_{\text{amb}}) + N_h\phi_\text{h,PHE}(T_\text{amb})}{ \phi_\text{s,PHE}(T_\text{amb})}
\end{split}
\end{equation}

The subscript PHE indicates that we are setting the interaction of Phenylalanine amino acid with water at ambient temperature, $\epsilon_{_\text{PHE},w}$. Now,  $\phi_\text{h,PHE}(T_\text{amb})=\epsilon_{_\text{PHE},w}$. Since $\mu_{_\text{LCW}}$ can be obtained from theory \cite{Huang2000}, we can now obtain $\omega_s$ by combining eqns. \eqref{eq:SurfaceHydrated} and \eqref{eq:LCWFit}.

We compare the results of the fitted parameters to the case $\alpha_s = \alpha_h = 0$. Here, the Hamiltonian corresponds to the temperature-independent potential fitted in Ref.~\cite{Abeln2011} with the adjustment that $\omega_s=1$ was used there.

Real amino acids vary in volume from 72 to 240 $\text{\AA}^3$~\cite{Mishra1984}. 
The above derivation assumes amino acids with a 3.3\AA\ radius, corresponding
to a volume of 150 $\text{\AA}^3$, We also derive the potential for amino acids with a size of 
225 $\text{\AA}^3$, corresponding to a radius that is 15\% larger. This yields the following values: $\alpha_s=4.5$ and $\alpha_h=11.5$.

\subsection{Definition of folded state}

Since in our simulations the folded ensemble is somewhat fluid and can contain configurations which are close to native, we need a way to define the folded ensemble. For this, we determined the point where the free energy with respect to the number of native contacts becomes significantly higher than zero. This was based on multiple free energy profiles. A representative free energy profile is shown in Figure \ref{fig:CutOffNint}.

\subsection{Heat Capacity of the system}


Validation of the model is done by determining the heat capacity of the system through its definition, $C_V = \frac{dQ}{dT}$, which can be evaluated in this model through
\begin{equation}
C_V=\frac{d\langle E \rangle}{dT},
\label{eq:CV}
\end{equation}
provided that no work is done on the system. In order to calculate $\langle
E \rangle$ we need to separate the hydrophobic potential,
$F_{\text{hydr}}$, into an enthalpic and entropic contribution. This can
be calculated using eqn. \eqref{eq:Fhydr} and the identity:

\begin{equation}
\langle E\rangle = \frac{ d\beta F}{d \beta}
\end{equation}
Note that $\langle E \rangle =\langle E_{\text{base}} \rangle$ for a temperature-independent potential. For the temperature-dependent potential this expression evaluates to $\langle E \rangle = \langle E_{\text{int}} \rangle+ \langle E_{\text{hydr}} \rangle $. At a fixed number of hydrophobe-water contacts we can calculate $\langle E_{\text{hydr}} \rangle$ analytically. 


Since $F_{\text{hydr}}$ is the only temperature-dependent part of the Hamiltonian $\mathcal{H}(T,\vec{r})$ defined in eqn. \eqref{eq:Hamiltonian}, $\langle E \rangle=\langle E_{\text{base}} \rangle+\langle E_{\text{hydr}}\rangle$, which can then be used to numerically estimate the derivative $\frac{d \langle E \rangle}{dT}$. The results are shown as black lines in Figure 3.

\subsection{Comparison heat capacity to experiments}

The heat capacity in the simulations are calculated at a constant volume ($C_V$), while the heat capacity is generally measured at constant pressure in experiments ($C_P$). While for gases the heat capacity can vary significantly, for water this difference is generally negligible. This can be shown by considering the identity for the difference in heat capacity at constant pressure and constant volume \cite{Schroeder2000}:

\begin{equation}
C_P-C_V = VT \frac{\alpha^2}{\beta_T}
\end{equation}

Where $\alpha$ is the thermal expansion coefficient, and $B_T$ is the isothermal compressibility, $V$ is the volume, and $T$ is the the temperature in K. For water, $\alpha=69\cdot10^{-6} \text{K}^{-1}$, and  $\beta_T = 4.6 \cdot 10^{-10} \text{m}^2 \text{N}^{-1}$. For water the difference between the heat capacity at constant pressure and the heat capacity at constant volume is approximately 0.6  \% at 293 K. The presence of a diluted protein does not significantly affect the isothermal compressibility and the thermal expansion coefficient. Therefore, the difference between the heat capacity at constant volume and the heat capacity at constant pressure is not significantly affected by the presence of diluted protein, which means that that we can directly compare the experimental and simulated heat capacities.

\subsection{Obtaining hydrophobic surface area}

We use the DSSP program \cite{Kabsch1983} to determine the surface accessible area of each residue for PDB-structures. We then define the following amino acids to be hydrophobic, in accordance with the amino acids that are considered hydrophobic in the lattice model:
Alanine, phenylalanine, cysteine, leucine, isoleucine, tryptophan, valine, methione, and tyrosine.
The sum of these contributions is shown on the y-axis in Figure 3C. The slope of the heat capacity is taken from \cite{Privalov2007} and multiplied by the sequence length found in the corresponding Protein-Database structure.

\subsection{Design procedure}

Because of the level of coarse graining present in the model, and the simplifying assumptions made in the potential, existing proteins will not fold to a correct structure. To find a sequence that will fold into a structure, it has to be specifically designed. The design procedure introduced by Shakhnovich and Gutin \cite{Shakhnovich1993a,Shakhnovich1993b,Shakhnovich1994} and used with small adaptations in Refs. \cite{Coluzza2003,Abeln2008,Abeln2011} uses a combination of the minimization of the potential energy of the folded structure and a measure to keep the variance of the amino acids high using a Monte Carlo algorithm:
\begin{equation}
P_{\text{Acc1}}=\min(1,\exp(-\Delta E_{\mathrm{base}} \beta_2))
\end{equation}
where $P_{\text{Acc1}}$ is the energy acceptance probability, $E_{\mathrm{base}}$ is the base pair potential energy of the folded state and $\beta_2$ is 1 divided by the design temperature.

We use a slightly adapted method where the potential energy difference between the native state and the fully extended state is minimized. This yields structures with a slightly higher heat denaturation temperature and a larger contribution of the hydrophobic interactions to the stability.

\subsection*{Protein sequence}

The protein sequence containing 80 amino acids was designed for a structure with 97 native contacts, and has a clear hydrophobic core.

\begin{texshade}{Sequence.fa}
\hideconsensus
\end{texshade}

\subsection{Discussion of pair potential}

The pair-potential in our paper, $\epsilon_{a(i),a(j)}$ is calculated with the same procedure as Hobohm \& Sander, but uses a larger dataset (see Ref \cite{Abeln2011}) based on PDB-select-25 \cite{Griep2010}. Due to the discrete nature of the lattice model long range interactions are not properly accounted for in the model. Since charged interactions are important for flexible and natively disordered proteins, this potential should not be used to model these proteins. Moreover, since flexible structures are harder to crystallize, the data contains a bias to soluble, rigid proteins. Nonetheless, a comparison potential derived from NMR-structures, where this bias is less pronounced, shows no significant differences in amino acid interactions derived from X-ray structures \cite{vanDijk2015}.

The temperature dependence of the free energy of transfer from the core of the protein to the solvent for hydrophobic residues is derived from hydrophobic theory. In previous work studying temperature dependent effects, this temperature dependence has been estimated from laboratory measurements of transfer \cite{Privalov1993} from a gaseous environment to water \cite{Wuttke2014}. While this approach provides a separate temperature dependence for each amino acid, the gas state may not capture the internal environment of the folded protein accurately. Nonetheless, for the hydrophobic residues this procedure yields values that are qualitatively similar to the results we obtain by using a statistical pair potential as a reference point and adding a temperature dependent term to the potential. A third approach, deriving a statistical pair potential from a large set of NMR-structures determined at differing temperatures, also yields qualitatively similar results. The quantitative differences between the potentials derived by the three methods can be explained by the differences in methodology.

The potential used is included in the following file:\\

Potential\_Data.txt\\

The data represents a two-dimensional array, where the order of the columns is the same as the order in the rows.

\setcounter{figure}{0}
\makeatletter
\renewcommand{\thefigure}{S\@arabic\c@figure}
\makeatother

\subsection{Umbrella sampling}
This simple model
allows for sufficient sampling of the conformational space at high
temperatures with respect to the number of native contacts using simple
Monte Carlo sampling. Umbrella sampling with the following quadratic biasing
potential was performed to verify the results obtained in this fashion:
\begin{equation}
E_{\text{umbr}}=\mathcal{H}(T,\vec{r}) + k(N_{\text{int}}-N_0)^2,
\end{equation}
where $\mathcal{H}(T,\vec{r})$ is the normal Hamiltonian, $N_{\text{int}}$ is the number of native contacts $k$ is the spring constant, and $N_0$ is the centre of the biasing potential. The spring constant $k$ was set to 0.02, and $N_0$ was set to values between 0 and 100, with a stepsize of 5, so $N_0 \in  \{0,5,10,15,...,95,100\}$. WHAM \cite{Grossfield2003} was then used to reconstruct the free energy surface with respect to $N_{\text{int}}$. We define a protein to be in the folded state when $N_{\text{int}} > 75$. The maximum number of native contacts is 97.

To explore the landscape of the native contacts together with the internal contacts $C_{\text{int}}$, umbrella sampling in two dimensions was performed. The potential used in this case was:

\begin{equation}
E_{\text{umbr}}=\mathcal{H}(T,\vec{r}) + k((N_{\text{int}} -N_0)^2 + (C_{\text{int}}-C_0)^2),
\end{equation}
where all values were identical to the previous simulations. The simulations were run with values for $C_0$ ranging from 0 to 100 internal contacts, with an interval of 5, so $C_0 \in  \{0,5,10,15,...,95,100\}$. Again, WHAM was used to reconstruct the free energy surface with respect to $C_{\text{int}}$. Note that while it is possible to have more than 97 internal contacts, this was not encountered often in our simulations and those configurations therefore had a high free energy. 
\subsection{Derivation Heat Capacity}

The heat capacity is given by $C_V=\frac{d \langle E \rangle}{dT}$ where $\langle E \rangle=\frac{d\beta F}{d\beta}$ is the expected value of the internal energy of the system. For a given configuration, the free energy $F$ equals the Hamiltonian $\mathcal{H}(T,\vec{r})$. We can use this to find we can find the expected potential energy of the system, $\langle E \rangle$ for a configuration.

\begin{equation}
F=E_{\text{base}}+F_{\text{hydr}}
\end{equation}

Substituting $F_{\text{hydr}}$ using eqn. \eqref{eq:Fhydr}

\begin{equation}
F=E_{\text{base}}- \alpha_s  N_{\text{s}}(T-T_{0,s})^2 -\alpha_h N_{\text{h}} (T-T_{0,h})^2
\end{equation}
Multiply both sides by $\beta$.
\begin{equation}
\begin{split}
\beta F = \beta E_{\text{base}} - \alpha_s N_s ( \frac{1}{\beta} - 2 T_{0,s} +\beta T_{0,s}^2) -\\
 \alpha_h N_{h} ( \frac{1}{\beta} - 2 T_{0,h} +\beta T_{0,h}^2)
\end{split}
\end{equation}

Taking derivative with respect to $\beta$ and using that $\frac{d\beta F} {d\beta} = \langle E \rangle$

\begin{equation}
\langle E \rangle = E_{\text{base}} - \alpha_s N_s ( T_{0,s}^2 -T^2) - \alpha_h N_h ( T_{0,h}^2 -T^2)
\end{equation}
$N_{\text{s}}$ represents the number of residues on the surface, while $N_{\text{h}}$ represents the number of residues that are fully solvated. The average over the configurational ensemble can subsequently be determined through Monte Carlo sampling, and the heat capacity can be determined through a numerical evaluation of eqn. \eqref{eq:CV}.

Assuming the folded structure consists of a single configuration, the heat capacity of the folded structure can be determined analytically:

\begin{equation}
C_V (T)=\frac{d\langle E \rangle }{dT}=2\alpha_s N_s T +2\alpha_h N_h T
\end{equation}

\begin{table*}[!p]
\subsection{Table of proteins}
\centerline{
\begin{tabular}{ l l l l }
\hline
Protein& PDB & $\frac{dC_P}{dT}$& Hydrophobic surface area \\ \hline
BPTI & 5pti & 37.7 & 1194 \\
Barnase & 1rnb & 76.3 & 1281 \\
Myoglobin& 1mbo& 122.4 & 1581 \\
Lysosyme & 1lzl & 45.1 & 754\\
Cytochrome C& 5cyt & 92.7 & 1279\\
Ubiquitin& 1ubq & 68.4 & 719 \\
$T_4$ lysozyme& 3lzm & 147.6 & 1814\\
RNase T1& 8rnt & 95.68 & 1074 \\
RNase A& 7rsa & 136.3 & 1193\\
Engrailed & 3hdd & 71.5 & 754 \\
Mat $\alpha$ 2& 1apl & 84.37 & 882 \\
Antennapedia& 9ant &82.88 &1012 \\
HGMD-74& 1hma & 112.42 & 1068\\
LZ-GCN4& 2zta & 54.25 & 554 \\
HMG SOX5 & 1I11 & 131.6 & 1536 \\
Zn-finger TFIIIA& 1tf3 & 159.16 & 1894 \\
NHP6A& 1cg7 & 192.51 & 1887 \\
SRY& 1j46 & 233.92 & 2335 \\
Lef-79& 2lef & 225.25 & 1876 \\
\hline
\end{tabular} }
\label{tab:ProteinTable}
Lists the PDB-ids, the slope of the heat capacity and the hydrophobic surface area for the proteins investigated.
\end{table*}

\subsection{Length scale dependence of solvation}
\begin{figure*}[p!]
\includegraphics{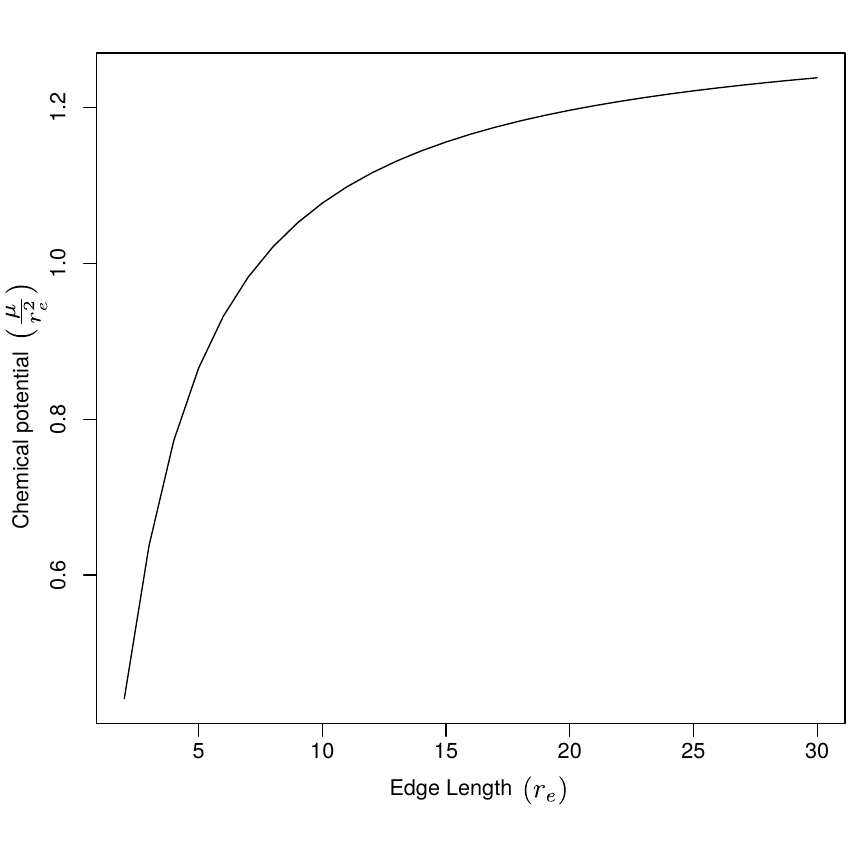}
\caption{ The length-scale dependence of the excess chemical potential of solvation, $\mu$, as a function of the edge length of a hydrophobic cube, $r_e$,  in the model. The values of the chemical potential are shown for $2\leq r_e \leq 30$. Theoretically \cite{Huang2000,Chandler2005}, the chemical potential scales with the volume of a solvated hydrophobe for small solutes, and with the surface for large solutes. A similar scaling is found in our model for both small and large solutes. }
\label{fig:Length_Normalized}
\end{figure*}

The excess chemical potential of solvation, $\mu$, shown in Figure~\ref{fig:Length_Normalized} is calculated by considering a hydrophobic homopolymer configured in a cubic formation of increasing dimensions. If the temperature is fixed, $\mu$ can be written as the sum of the residue-solvent interactions:

\begin{equation}
\mu =  r_s \epsilon_{s,w}+  r_h \epsilon_{h,w}
\end{equation}
\noindent
Where $r_s$ the number of residues at the surface, $\epsilon_{s,w}$ is the interaction energy for a surface residue with the solvent, and $\epsilon_{h,w}$ is the interaction of a fully solvated residue with the solvent. In our model, $\epsilon_{s,w} = \omega_s \epsilon_h$. As described in the methods, the weight $\omega_s$ is fitted to  0.32. Note that buried residues do not contribute to the transfer chemical potential since they are not in contact with the solvent.

To compare the relation of the chemical potential with the size of the hydrophobic cluster in our model, we consider a cube of different dimensions. Let $r_e$ be the number of residues along an edge. Except for the special case $r_e=1$, there are no fully solvated residues. That means that the chemical potential is:
\begin{equation}
\mu = r_s \epsilon_{s,w}
\end{equation}
Therefore, computing the number of surface residues allows us to find the chemical potential. This is achieved by subtracting the number of residues that are buried, $(r_e-2)^3$, from the total number of residues, $r_e^3$. 

\begin{equation}
r_s =  r_e^3-(r_e-2)^3 = 6 r_e^2-12 r_e + 8
\end{equation}

This means that the chemical potential can be written as:

\begin{equation}
\mu =  \epsilon_{s,w} ( 6 r_e^2-12 r_e + 8)
\end{equation}

The result of this equation is plotted in Figure \ref{fig:Length_Normalized}.

\subsection{Simplified model}

To test the influence of the length scale dependence on the results, we have included simulations with a simplified model that does not incorporate this length scale dependence, by setting $\alpha_s = \alpha_h$. This results in qualitatively similar results, as shown in Figures \ref{fig:EnergyLandscape} and \ref{fig:HeatCapacity1}. 

\begin{figure*}[p!]
\includegraphics{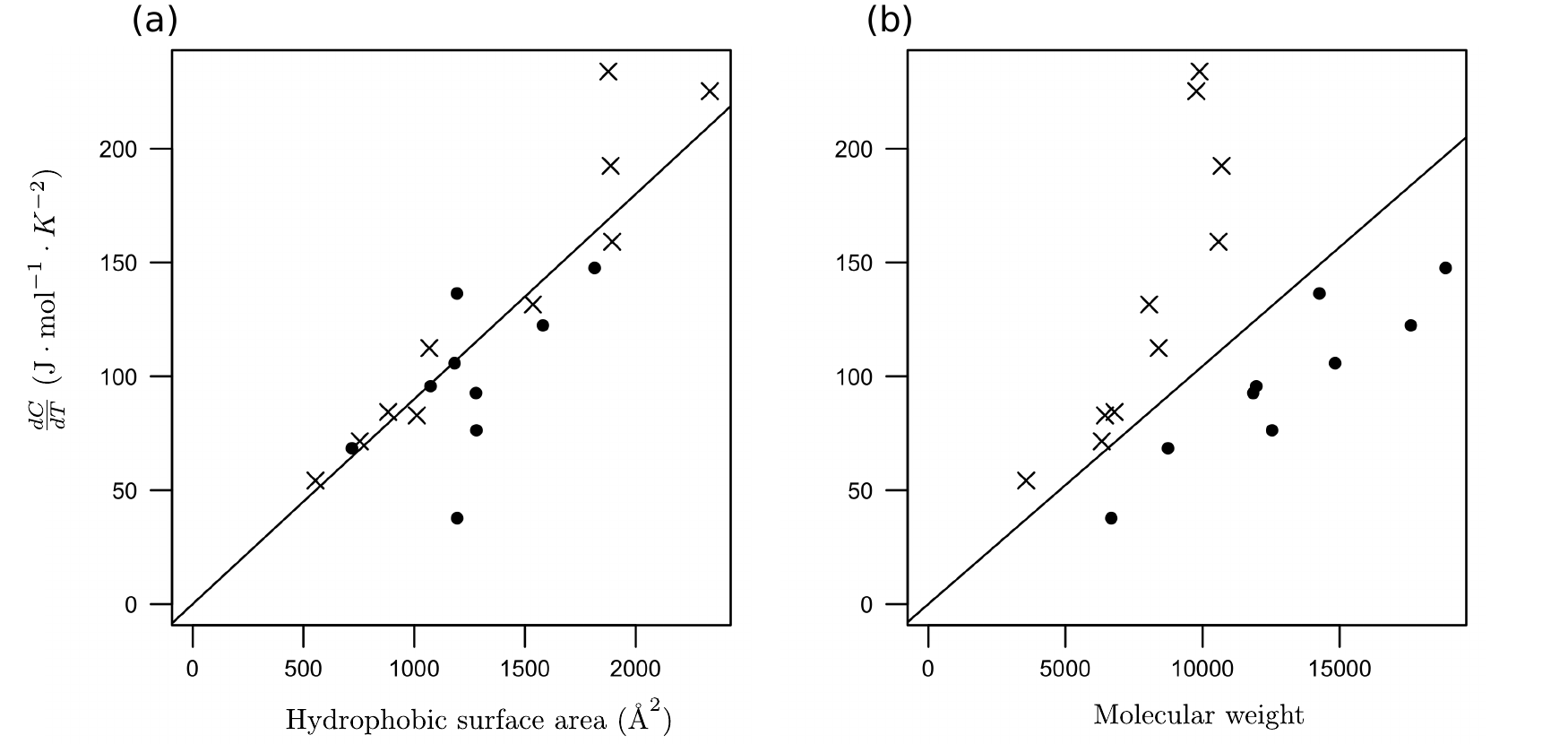}
\caption{Hydrophobic surface area vs Slope of heat capacity (a) and Molecular weight vs Slope of heat capacity (b).
A similar analysis in \cite{Naganathan2011} for an extended set of proteins, shows the same pattern but does not recognize that the hydrophobic surface area provides a fit that is valid for both normal globular proteins (dots) and DNA-binding proteins (crosses).}
\label{fig:Slope_bound}
\end{figure*}

\begin{figure*}[p!]
\includegraphics{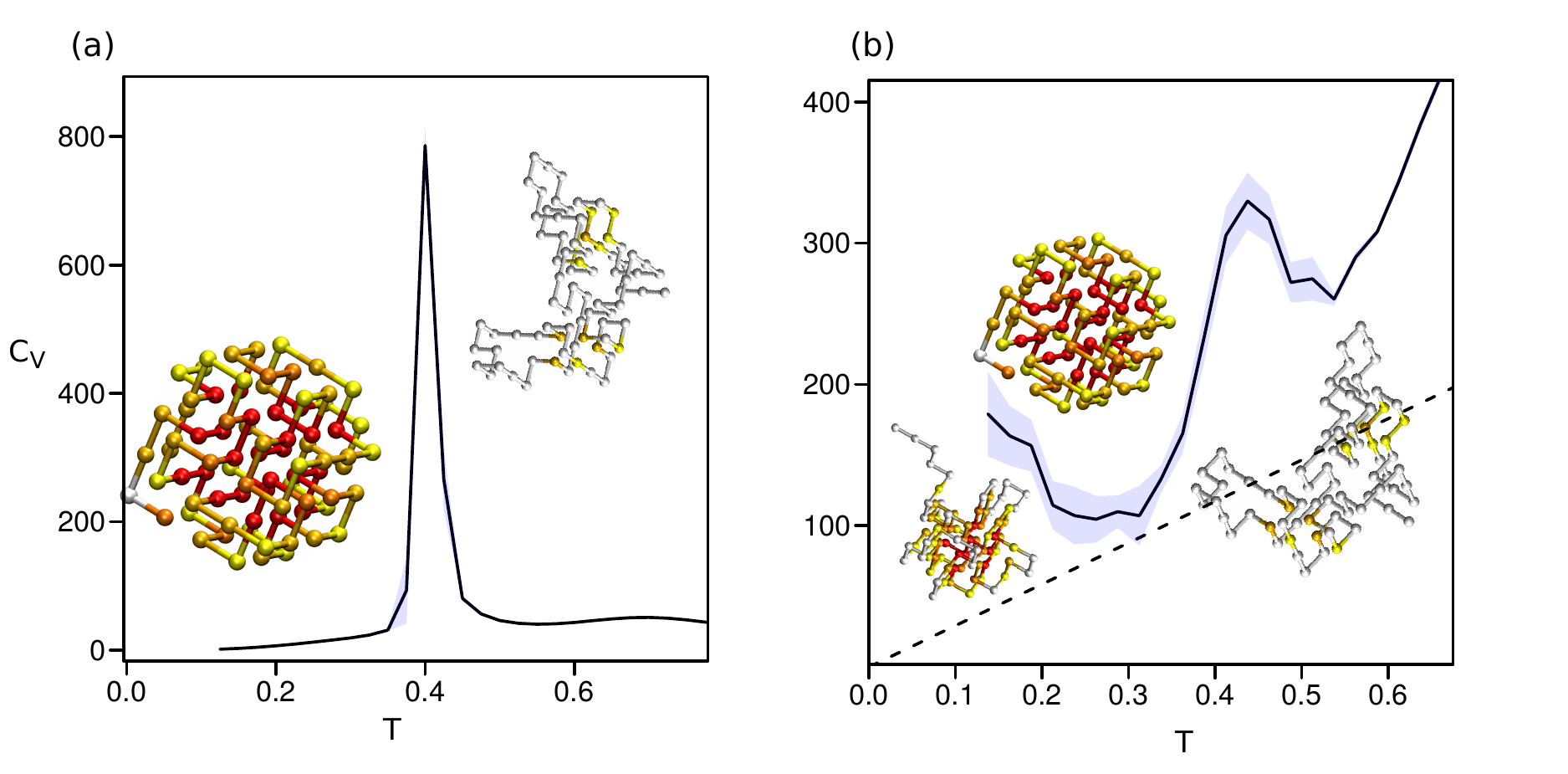}
\caption{Heat capacity for temperature independent potential (a) and temperature dependent potential (b). (a) Heat capacity for a temperature independent potential, similar to earlier published results, we find a single peak for the heat induced denaturation. (b) Heat capacity versus temperature for bigger amino acids. A double peak can be seen for the heat capacity, indicating two phase transitions.}
\label{fig:DoublePeak}
\end{figure*}

\begin{figure*}[p!]
\includegraphics{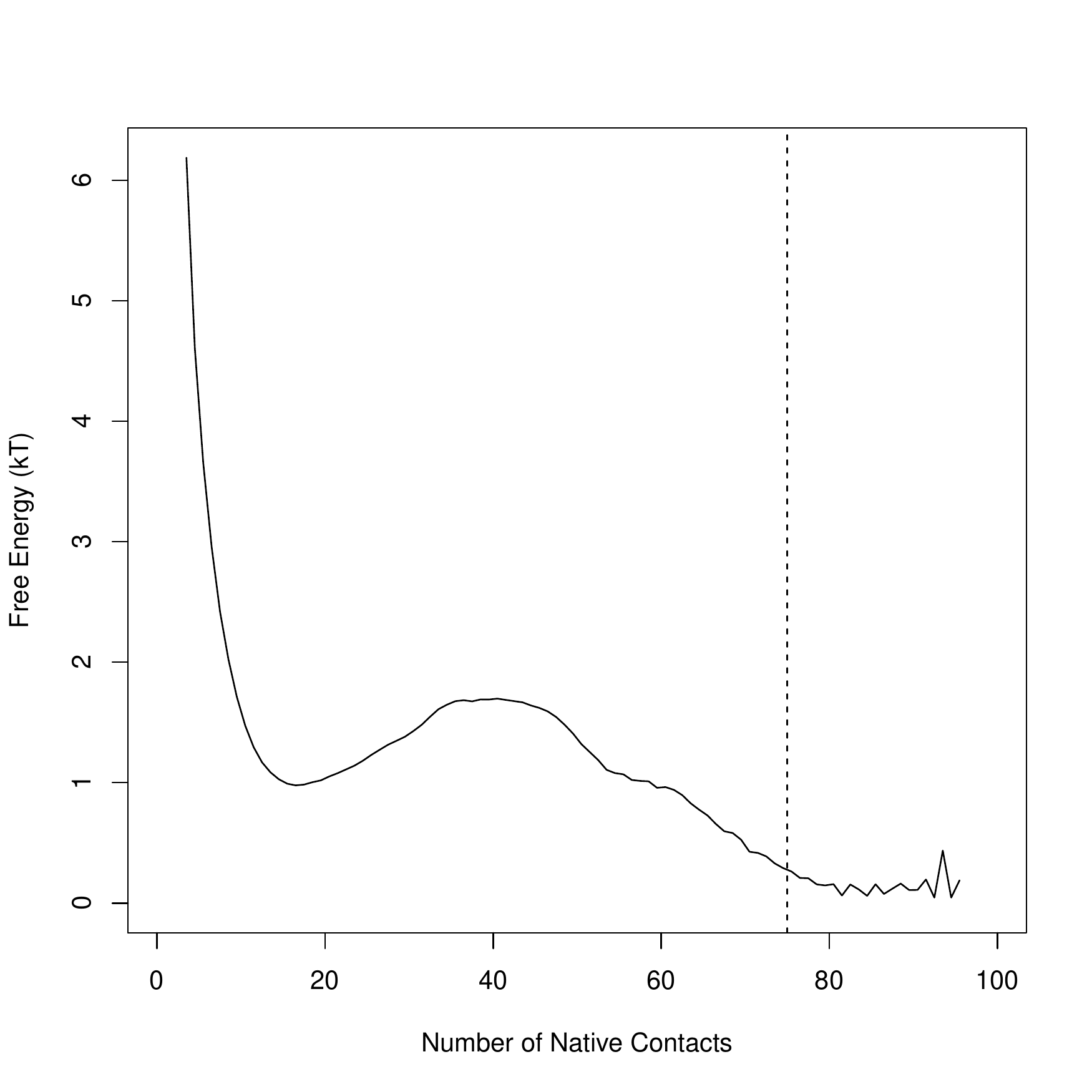}
\caption{Free Energy landscape with respect to number of native contacts for the temperature dependent potential at $T=0.4$. Since in our simulations the folded ensemble is somewhat fluid and can contain configurations which are close to native, we need a way to define the folded ensemble. Moreover, there are some artifacts near the native state in the energy landscape due to the discrete nature of the lattice model. The definition of the folded state was determined as the point where the free energy with respect to the number of native contacts becomes significantly higher when compared to the minimum.}
\label{fig:CutOffNint}
\end{figure*}

\begin{figure*}
  \centering
  \includegraphics{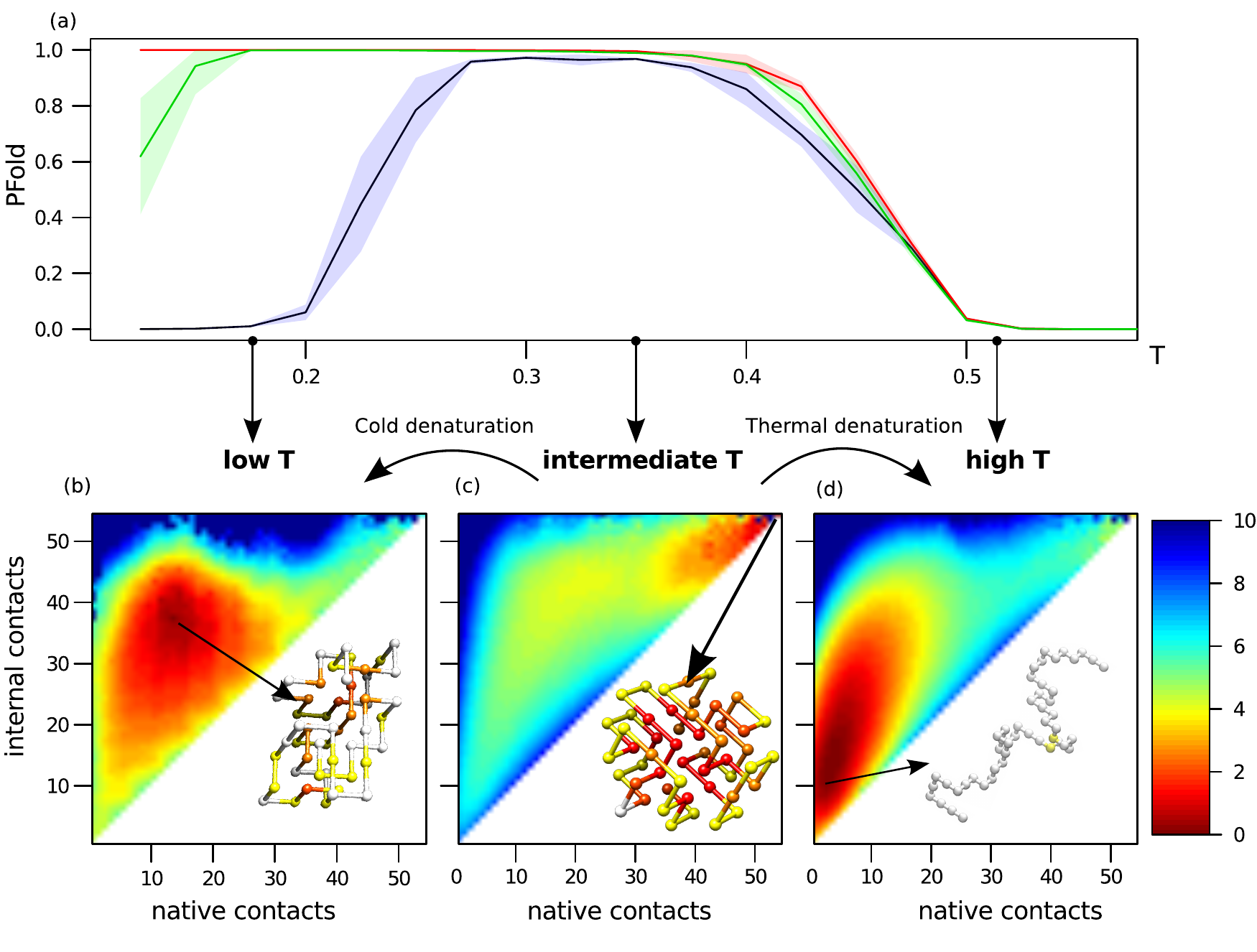}
  \caption{Temperature-dependent folding stability and structure.\\ (a)
    The probability for the model protein to be in the folded state, $P_{\text{Fold}}$, versus
    temperature, with $\alpha_s=\alpha_h=0$ (red), $\alpha_s=\alpha_h=5 k_B T$ (green) and $\alpha_s=\alpha_h=10 k_B T$
    (black). A protein is considered folded if more than~$45$ of the
    $54$~possible native contacts are present. The 95\% confidence interval is indicated by the shaded area.\\
    (b)--(d) Free energy landscapes for the number of native contacts
    ($N_{\text{int}}$) and all internal contacts ($C_{\text{int}}$) for
    $\alpha_s=\alpha_h=10 k_B T$ at (b) low temperature ($T=0.175$), (c) intermediate
    temperature ($T=0.375$), and (d) high temperature ($T=0.575$).  For
    $\alpha_s = \alpha_h=10 k_B T$ the protein denatures at low temperatures, with many exposed
    hydrophobic amino acids. However, this denatured structure is a lot more
    compact than the heat denatured protein, and there are less native contacts present. At intermediate temperatures
    the protein has the highest stability in its folded configuration
    (indicated by the arrow) where $N_{\text{int}}=C_{\text{int}}=54$. At
    high temperatures the protein makes only transient contacts.}
  \label{fig:EnergyLandscape}
\end{figure*}

 \begin{figure*}
\includegraphics{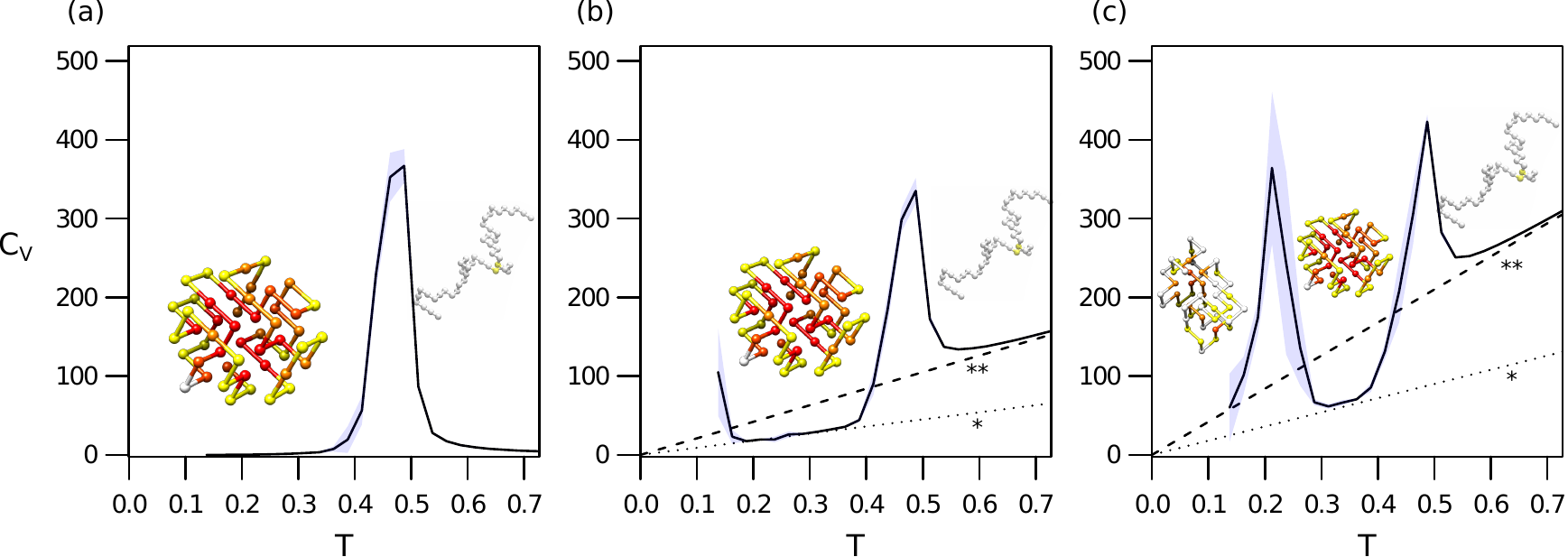}
\caption{Heat capacity versus temperature for $\alpha_s=\alpha_h =0$ (a), $\alpha_s = \alpha_h=5 k_B T$ (b) and $\alpha_s = \alpha_h=10 k_B T$ (c).\\ 95 \% confidence region is shown through shading. The dotted lines indicate the baselines of the native state (*) and the denatured state (**), obtained through a simple calculation (see section ``Derivation Heat Capacity'').\\
 (a) The classic model shows a single peak in the heat capacity at the folding transition. (b) The heat capacity for $\alpha_s = \alpha_h=5 k_B T$, as calculated by $C_V=\frac{dE}{dT}$ does not show cold denaturation, but does show a linear increase in the heat capacity with respect to temperature. (c) The heat capacity clearly shows two sharp peaks signifying two phase transitions. The linear increase in heat capacity in the absence of a phase transition is also observed experimentally.}

\label{fig:HeatCapacity1}
\end{figure*}

\bibliography{MyCollection08212014}

\end{document}